\documentclass[aps,epsf,10pt,showpacs,floatfix,showkeys]{revtex4}
\usepackage{graphicx}
\usepackage{mathrsfs}
\usepackage{amsmath}
\usepackage{amsfonts}
\usepackage{amssymb}
\usepackage{epsfig}
\usepackage{mathrsfs}                                                                                

\usepackage{theorem}
\theorembodyfont{\upshape}

\newcommand{\be}{\begin{equation}}
\newcommand{\ee}{\end{equation}}
\newcommand{\bea}{\begin{eqnarray}}
\newcommand{\eea}{\end{eqnarray}}
\newcommand{\nn}{\nonumber \\}

\input epsf.sty
\usepackage{epsf}
\newlength{\diaght}
\begin{document}

\author{I. Lyberg}
\email{ivar.lyberg@uclouvain.be}
\affiliation{Unit\'e de physique th\'eorique et math\'ematique, Universit\'e catholique de Louvain, 1348 Louvain-La-Neuve, Belgium}
\date{\today}
\title{The Ising lattice with Brascamp-Kunz boundary conditions and an external magnetic field}
\begin{abstract}The partition function of the finite and non-isotropic Ising lattice with Brascamp-Kunz boundary conditions is calculated exactly in the absence of an external magnetic field and for an external magnetic field $H/kT=i\pi/2$. 
\end{abstract}

\keywords{Ising lattice, partition function, Brascamp-Kunz boundary conditions.}
\pacs{05.50.+q, 02.30.Ik, 75.10.Pq.}

\maketitle

\section{Introduction}
\label{int}
Nickel {\cite{ni} \cite{ni2}} found singularities in the susceptibility of the two dimensional isotropic Ising lattice in the absense of an external magnetic field. Explicitly, he found that for the isotropic Ising lattice with lattice constant $K$, the susceptibility  has a natural boundary on the circle 
\be |\sinh{2K}|=1,
\label{nb}
\ee
unlike the free energy and the magnetization. That is, the set of singularities of the susceptibility is dense on the circle (\ref{nb}), whereas the sets of singularities of the free energy and the magnetization are not; in fact, they are dense only at isolated points. The reason for this is not yet known, but there are som suggestive findings regarding differences between the susceptibility on the one hand and the free energy and the magnetization on the other. A {\it differentiably finite} function is one that is the solution of a linear ordinary differential equation of finite order with polynomial coefficients. Guttmann and Enting {\cite{guttmann}} found that while (in the absence of an external magnetic field) the partition function and the magnetization are both differentiably finite functions, there is strong evidence that the susceptibility is not. They also found that the susceptibility most likely has a natural boundary even in the non-isotropic case. Specifically, what they found was that for an anisotropic lattice with lattice constants $K_1$ and $K_2$, the susceptibility $\chi(z_1,z_2)$, where
\be z_l:=\tanh{K_l}~(l=1,2),
\label{zl0}
\ee
has a natural on the circle $|z_1|=1$ for fixed $z_2$. Further evidence supporting this claim was given by Orrick, Nickel, Guttmann and Perk \cite{or}.

The Hamiltonian of the Ising lattice of size $(M,N)$ in an external magnetic field $H$ is 
\bea \mathcal{E}:=-E_1\sum_{j=1}^M\sum_{k=1}^N\sigma_{j,k}\sigma_{j,k+1}-E_2\sum_{j=1}^{M+1}\sum_{k=1}^N\sigma_{j,k}\sigma_{j+1,k}-H\sum_{j=1}^{M}\sum_{k=1}^N\sigma_{j,k},
\label{hi0}
\eea
where no boundary conditions have been specified. Denote by $Z_{M,N}$ the corresponding partition function. Exact solutions may be found for the particular magnetic fields $H= 0$ and $H/kT= i\pi/2$. In what follows, the dimensionless parameters
$K_l:= E_l/kT~(l=1,~2)$ and $h:=H/kT$
will be used instead of $E_1$, $E_2$ and $H$. It is customary to use the variables 
\be x_l:=e^{-2K_l}~(l=1,2)
\label{xl}
\ee 
and 
\be z:=e^{-2h} 
\label{zdef}
\ee
instead of $K_1$, $K_2$ and $h$. 

It was conjectured by Fisher {\cite{f}} that in the isotropic case $K:=K_1=K_2$, the zeros of $Z_{M,N}$ in $x:=e^{-2K}$ will all approach the circles 
\be |x\pm 1|=\sqrt{2},
\label{xpm1}
\ee 
or equivalently the circle (\ref{nb}), in the thermodynamic limit $M,N\rightarrow \infty$. Lee and Yang {\cite{ly}} showed that if $0\leq x^2 \leq 1$, then zeros of $Z_{M,N}$ in $z$ all lie on the unit circle $|z|=1$. Wood {\cite{wood}} showed that in the non-isotropic case $(K_1,K_2)=(K,\alpha K)$ ($\alpha\in {\bf R}$), the zeros of the partition function will all approach the curve described by the polar equation
\be \cos{\theta}=-\frac{r^{2\alpha}-1}{r^{2\alpha}+1}\frac{1+r^2}{2r}
\label{polar}
\ee
in the thermodynamic limit, where $\theta$ is the polar angle. In particular, this means that the zeroes of the partition function in the anisotropic case with lattice constants $K_1$ and $K_2$ will spread over a two dimensional region (as $\alpha$ is varied). It should thus be possible to analytically continue the free energy through this region. The transformation $K_2\rightarrow -K_2$ reflects the curve (\ref{polar}) about the imaginary $x$ axis. Thus, in particular, for $\alpha=1$ these two curves are the two circles (\ref{xpm1}).

In the thermodynamic limit $M,N\rightarrow\infty$, the free energy
\be f(z,x_1,x_2):=kT\lim_{M,N\rightarrow\infty}\frac{1}{MN}\log{Z_{M,N}} 
\ee
will only depend on $z$, $x_1$ and $x_2$. Away from the zeros of the partition function, $f$ will be an analytic function of $x_1$, $x_2$ and $z$. 

\subsection{The Brascamp-Kunz lattice}
\begin{figure}[bt]
{\epsfig{file=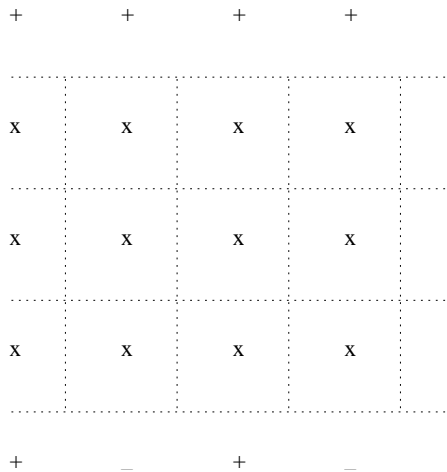, width=6cm}}
\caption{The lattice $\Lambda$ (marked by ``x''; $M=3$, $N=4$) and the dual lattice $\Lambda^{*}$ (at intersections of lines).}
\label{fig0}
\end{figure}

The partition function $Z_{M,N}$ of a finite lattice with the Hamiltonian (\ref{hi0}) has been calculated exactly for many different boundary conditions, including for the Ising lattice on a M\"obius strip and on a Klein bottle {\cite{mobius}}. However, the boundary conditions particularly useful for studying the natural boundary are the so called Brascamp and Kunz boundary conditions, described in section \ref{results}. Brascamp and Kunz {\cite{BK}}, using a result by McCoy and Wu {\cite{MW}}, showed that for the Ising lattice with so called Brascamp-Kunz boundary conditions, the zeros do not just approach the set (\ref{xpm1}) asymptotically, but lie precisely in it. Brascamp and Kunz constructed the boundary conditions named after them from McCoy's and Wu's calculation of the partition function of a cylindrical lattice with a magnetic field $\mathfrak{H}$ on the lower boundary. McCoy and Wu considered a general $\mathfrak{H}$ (although strictly speaking $\mathfrak{H}/kT\neq i\pi/2$), but Brascamp and Kunz found that in the limit $\mathfrak{H}/kT \rightarrow i\pi/2$, the expression for the partition function much simplifies, and the zeros of the partition function all lie on the circle (\ref{nb}). They then used the duality relations presented in ref. \cite{B} to construct a lattice whose dual lattice is the lattice of McCoy and Wu with $\mathfrak{H}/kT=i\pi/2$. This lattice is the Brascamp and Kunz lattice. Specifically, it is the two dimensional cylindrical Ising lattice $\Lambda=\{(m,n)~|~1\leq m \leq M,~1\leq n \leq N,~(m,N+1)=(m,1)\}$ with the following boundary conditions.\newline
{\it (i)} The lattice interacts with a row of fixed, positive spins above it,\newline 
{\it (ii)} The lattice interacts with a row of fixed, alternating spins below it.\newline 
(See fig. \ref{fig0}). The interaction between neighboring spins on $\Lambda$ is $E_1$ in the horizontal direction and $E_2$ in the vertical direction. Apart from from interactions between nearest neighbors, there is an external magnetic field $H(j,k)$. In this paper, $H(j,k)$ will constant; either $H(j,k)=0$ for all $(j,k)$ or $H(j,k)/kT=i\pi/2$ for all $(j,k)$. 
Thus the Hamiltonian is defined as
\bea \mathcal{E}_{\Lambda}(\sigma,E_1,E_2,H)=-E_1\sum_{j=1}^M\sum_{k=1}^N\sigma_{j,k}\sigma_{j,k+1}-E_2\sum_{j=0}^M\sum_{k=1}^N\sigma_{j,k}\sigma_{j+1,k}-\sum_{j=1}^M\sum_{k=1}^NH(j,k)\sigma_{j,k}
\label{hi}
\eea
where $\sigma_{j,k}=\sigma_{j,k+N}=\pm 1$, $\sigma_{0,k}=1$ and $\sigma_{M+1,k}=(-1)^{k+1}$.

\subsection{The magnetic field $H/kT=i\pi/2$}

The two dimensional isotropic Ising lattice in the presence of an external magnetic field $H/kT=i\pi/2$ was first studied by Lee and Yang {\cite{ly}}. Since then, this problem has been much studied, both for isotropic {\cite{g} \cite{wu} \cite{ms2}} and anisotropic lattices{\cite{MW2}}. However, to the authour's knowledge, no study has been made of a lattice with Brascamp-Kunz boundary conditions in an external magnetic field.

As will be shown in section \ref{results}, the set of zeros of an isotropic lattice with these boundary conditions in a magnetic field $H/kT=i\pi/2$ will lie on the one dimensional, connected set; moreover, in the thermodynamic limit the set of zeros will become dense on this set. In this sense the magnetic fields $H= 0$ and $H/kT= i\pi/2$ are similar. It has been conjectured elsewhere {\cite{m}} that for any other magnetic field, the set of zeros does not become dense on a set that has even a one dimensional subset. Future studies may reveal the relationship between Nickel singularities and the particular magnetic field $H=0$. It is yet unclear whether there are Nickel singularities for $H\neq 0$, or only for $H=0$.

Lee and Yang {\cite{ly}} showed that the isotropic Ising lattice in a magnetic field $i\pi/2$ is equivalent to a two dimensional lattice gas. Let the free energy per site in the thermodynamic limit be $f$. Lee and Yang {\cite{ly}} conjectured that the magnetization $I(i\pi/2,K,K):=(kT)^{-1}(\partial f/\partial h)|_{h=i\pi/2}$ is 
\be I(i\pi/2,K,K)=\left(\frac{(1+x^2)^2}{1-x^2}(1+6x^2+x^4)^{-1/2}\right)^{1/4}
\label{ik}
\ee
where 
\be x:=e^{-2K}.
\label{x}
\ee
Later McCoy and Wu {\cite{MW2}} proved (\ref{ik}) by calculating $\lim_{n\rightarrow \infty}S_n=:I^2$ {\cite{mpw}}, where the row correlation function $S_n:=\langle\sigma_{0,0}\sigma_{0,n}\rangle$ is the expectation value of the product $\sigma_{0,0}\sigma_{0,n}$ of spins at sites $(0,0)$ and $(0,n)$. Moreover, they calculated the magnetization for the non-isotropic lattice with lattice constants $K_1$ and $K_2$ in the horizontal and vertical directions, respectively. They found
\be I(i\pi/2,K_1,K_2)=\left(\frac{1}{2}(z_1+z_1^{-1})(z_2+z_2^{-1})(z_1^{2}+z_1^{-2}+z_2^{2}+z_2^{-2})^{-1/2}\right)^{1/4}
\label{ini}
\ee
where 
\be z_l:=\tanh{K_l},~l=1,2.
\label{zld}
\ee
Gaaff {\cite{g}} found that the Ising lattice with an external magnetic field $H$ is in fact equivalent to a sixteen vertex model, which simplifies to an eight vertex model when $H/kT=i\pi/2$. Wu {\cite{wu}}, using the same approach as Gaaff, found that the Ising lattice has no phase transition in a magnetic field $H/kT=i\pi/2$, unless additional diagonal interactions and/or four-spin interactions are introduced. 

While numerical studies have been made on the problem of the Ising lattice in a general magnetic field {\cite{ms}}, there are no theorems on the analyticity of the free energy in this case. It should be possible to obtain the analytic properties of the free energy $f$ from the zeroes of the partition function $Z_{M,N}$. $f$ can only depend on $K_1$, $K_2$ and $h$. When $K_1=K_2=K$, one thus seeks the zeroes of $Z_{M,N}$ in the planes of $z$ and $x$ respectively. However, no systematic study of the zeros of the partition function has yet been made. The finding that the set of zeros becomes dense on a one dimensional set both for $H=0$ and $H/kT=i\pi/2$ may be useful for such a study.

\section{Summary of results} 
\label{results}
The partition function 
\be Z_{\Lambda}(K_1,K_2,h)=\sum_{\sigma \in \{-1,1\}^{\Lambda}}\exp{-\mathcal{E}_{\Lambda}(\sigma,E_1,E_2,H)/kT}
\label{pf}
\ee
will be calculated for the constant external magnetic fields $H\equiv 0$ and $H/kT\equiv i\pi/2$ for which the problem is exactly solvable. Brascamp and Kunz \cite{BK} calculated $Z_{\Lambda}(K,K,0)$. $Z_{\Lambda}(K_1,K_2,0)$ and $Z_{\Lambda}(K_1,K_2,i\pi/2)$ are given by
\begin{equation}
Z_{\Lambda}(K_1,K_2,0)=x_1^{-MN/2}x_2^{-MN/2}\prod_{j=1}^{N/2}\prod_{k=1}^{M}\{(1+x_1^2)(1+x_2^2)-2x_2(1-x_1^2)\cos{\theta_j}-2x_1(1-x_2^2)\cos{\varphi_k}\}\label{p}
\end{equation}
and
\bea Z_{\Lambda}(K_1,K_2,i\pi/2)&=&u_1^{-MN/4}u_2^{-MN/4}\nn
&&\prod_{j=1}^{N/2}\prod_{k=1}^{M/2}\{(1+u_1^2)(1+u_2^2)-4u_1u_2-2u_2(1-u_1)^2\cos{2\theta_j}-2u_1(1-u_2)^2\cos{\phi_k}\},\label{bk2}
\eea
where 
\bea &&\theta_j=(2j-1)\pi/N,~\varphi_k=k\pi/(M+1),~\phi_k=(2k-1)\pi/(M+1)\nn 
{\rm and}~&&u_l=e^{-4K_l}~(l=1,~2).
\label{def}
\eea
In particular the free energy per site in the thermodynamic limit, $f$, is 
\bea && f/kT:=\lim_{M,N\rightarrow \infty}\frac{1}{MN}\log{Z_{\Lambda}(K_1,K_2,i\pi/2)}=K_1+K_2+\frac{1}{(2\pi)^2}\int_{0}^{\pi}dx\int_{0}^{\pi}dy\nn
&&\log{\{(1+u_1^2)(1+u_2^2)-4u_1u_2+2u_1(1-u_2)^2+2u_2(1-u_1)^2-4u_2(1-u_1)^2\cos^2{x}-4u_1(1-u_2)^2\cos^2{y}\}},\label{bk2i}
\eea
which is the same as the result found by McCoy and Wu {\cite{MW2}}.

As has been already mentioned, with Brascamp-Kunz boundary conditions, the zeroes of $Z_{\Lambda}(K,K,0)$ lie on the circle (\ref{nb}) even when the lattice is finite, as can be seen from (\ref{p}). In the case $h\equiv i\pi/2$, the set of zeroes will also be one dimensional with Brascamp-Kunz boundary conditions in the special case $K_1=K_2$. When $K_1=K_2$, (\ref{bk2}) becomes 
\bea Z_{\Lambda}(K,K,i\pi/2)=(u^{-1}-1)^{MN/2}\prod_{j=1}^{N/2}\prod_{k=1}^{M/2}\{1+u^2+u(6-4\cos^2{\phi_k/2}-4\cos^2{\theta_j})\}.
\label{z}
\eea
The zeroes lie on the circle
\be |u|=1~{\rm for}~-1\leq 3-2\cos^2{\theta_j}-2\cos^2{\phi_k}\leq 1
\ee
and on the line segment
\be -3-2\sqrt{2}\leq u\leq -3+2\sqrt{2}~{\rm for}~1\leq 3-2\cos^2{\theta_j}-2\cos^2{\phi_k}\leq 3.
\ee

\section{The calculation of $Z_{\Lambda}(K_1,K_2,0)$}
\label{bk}
It will be shown that the partition function $Z_{\Lambda}(K_1,K_2,0)$ is given by (\ref{p}).

Brascamp and Kunz {\cite{BK}} used a result found by McCoy and Wu {\cite{MW}} to get their result. Ref. \cite{MW} concerns a lattice on a finite cylinder with a magnetic field on the lower boundary. To prove (\ref{p}), we consider the dual lattice $\Lambda^{*}=\{(m,n)~|~1\leq m \leq M+1,~1\leq n \leq N,~(m,N+1)=(m,1)\}$, shown in fig. \ref{fig0}: It is periodic in the horizontal direction and free along the upper boundary. The lattice constants are $K_2^*$ in the horizontal direction and $K_1^*$ in the vertical direction, where
\be \sinh{2K_1}\sinh{2K_1^*}=\sinh{2K_2}\sinh{2K_2^*}=1.
\label{dr0}
\ee
Therefore there may be defined a Hamiltonian on $\Lambda^{*}$ given by
\bea \mathcal{E}_{\Lambda^*}(\sigma,K_1^*,K_2^*,H^*):=-E_1^*\sum_{j=1}^M\sum_{k=1}^N\sigma_{j,k}\sigma_{j+1,k}-E_2^*\sum_{j=1}^{M+1}\sum_{k=1}^N\sigma_{j,k}\sigma_{j,k+1}-\sum_{j=1}^{M+1}\sum_{k=1}^NH^*(j,k)\sigma_{j,k},
\label{hdual}
\eea
and a corresponding partition function of $\Lambda^{*}$ given by
\be Z_{\Lambda^*}(K_1^*,K_2^*,h^*)=\sum_{\sigma \in \{-1,1\}^{\Lambda^*}}\exp{-\mathcal{E}_{\Lambda^*}(\sigma,K_1^*,K_2^*,H^*)/kT^*}.
\label{dpf}
\ee
In order to continue, it is necessary to establish the following lemma:

{\bf Lemma} 
 \bea Z_{\Lambda}(K_1,K_2,0)=2^{-1-N/2}(\sinh{2K_1})^{MN/2}(\sinh{2K_2})^{(M+1)N/2}Z_{\Lambda^{*}}(K_1^*,K_2^*,\mathfrak{h})
\label{Zb}
\eea
where $\mathfrak{h}$ is an external magnetic field on $\Lambda^*$ given by
\be \mathfrak{h}(m,n)=\frac{i\pi}{2}\delta_{m,M+1}.
\ee

\begin{figure}[bt]
{\epsfig{file=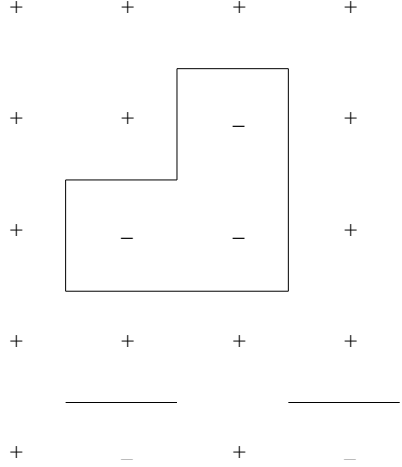, width=5.25cm}}
\caption{A spin configuration of the lattice $\Lambda$ with no external magnetic field. Here $M=3$ and $N=4$. A perpendicular line is drawn between any two spins of opposite signs \cite{B}. In this way, this particular configuration gives rise to one closed loop and two dimers on the dual lattice $\Lambda^{*}$. The latter give the appearance of a magnetic field $\mathfrak{H}^*/kT^*=i\pi/2$ on the lower boundary of $\Lambda^{*}$.}
\label{fig1}
\end{figure}

Consider fig. \ref{fig1}. There is a perpendicular line drawn between any two spin sites on $\Lambda$ whose spins have product $-1$. With this particular spin configuration, each site on the lower boundary of the dual lattice $\Lambda^*$ has precisely one dimer intersecting it. It is easy to see that with any spin configuration, each site on the lower boundary of $\Lambda^*$ will have one or three dimers intersecting it. Let a dimer carry the weight of the lattice constant on $\Lambda$ that it crosses, and let the collection of dimers corresponding to a spin configuration be called $\lambda$. The energy of a general configuration is 
\be E(\lambda)=-MNE_1-(M+1)NE_2+2|\lambda|
\label{el}
\ee
where $|\lambda|:=hE_2+vE_1$ and $h$ and $v$ are the number of horizontal and vertical dimers, respectively. In this way the partition function may be written as a ``low temperature'' expansion
\be Z_{\Lambda}(K_1,K_2,0)=e^{MNK_1+(M+1)NK_2}\sum_{\lambda}e^{-2|\lambda|/kT},
\label{zl}
\ee 
where the sum is over all possible paths $\lambda$.

Now consider the dual lattice $\Lambda^*$ on its own. $\Lambda^*$ is a cylindrical lattice with free boundary conditions. Suppose the magnetic field $\mathfrak{h}$ is applied. Since $\exp{\sigma i\pi/2}=i\sigma$, it follows from (\ref{hdual}) and (\ref{dpf}) that
\be Z_{\Lambda^*}(K_1^*,K_2^*,\mathfrak{h})=\sum_{\sigma \in \{-1,1\}^{\Lambda^*}}\left(\prod_{k=1}^{N}i\sigma_{M+1,k}\right)\exp{-\mathcal{E}_{\Lambda^*}(\sigma,K_1^*,K_2^*,0)}.
\label{dpf2}
\ee
Since $\exp{K_l^*\sigma\sigma'}=\cosh{K_l^*}+\sigma\sigma'\sinh{K_l^*}$, this can be written as a ``high temperature'' expansion
\bea && Z_{\Lambda^*}(K_1^*,K_2^*,\mathfrak{h})=(\cosh{K_1^*})^{MN}(\cosh{K_2^*})^{(M+1)N}\nn
&&\sum_{\sigma \in \{-1,1\}^{\Lambda^*}}\left(\prod_{k=1}^{N}i\sigma_{M+1,k}\right)\Bigg(\prod_{j=1}^{M}\prod_{k=1}^{N}(1+z_1^*\sigma_{jk}\sigma_{j+1,k})\Bigg)
\left(\prod_{m=1}^{M+1}\prod_{n=1}^{N}(1+z_2^*\sigma_{mn}\sigma_{m,n+1})\right),
\label{dpf3}
\eea
where $z_l^*:=\tanh{K_l^*}$. Rewrite the sum as 
\bea && Z_{\Lambda^*}(K_1^*,K_2^*,\mathfrak{h})=(\cosh{K_1^*})^{MN}(\cosh{K_2^*})^{(M+1)N}\nn
&&\sum_{\sigma \in \{-1,1\}^{\Lambda^*}}\left(\prod_{k=1}^{N/2}i^2z_2^*\sigma_{M+1,2k-1}\sigma_{M+1,2k}\right)
\left(\prod_{\{x,y\in \Lambda^*~|~|x-y|=1\}}(1+\kappa(x,y)\sigma_{x}\sigma_{y})\right)
,
\label{dpf3b}
\eea
where
\bea
\kappa(x_j,y_j) := \left\{ \begin{array}{ll}
 z_1^* & \textrm{if $x_j-y_j=(\pm 1,0)$},\\
 z_2^* & \textrm{if $x_j-y_j=(0,\pm 1)$ and $(x_j,y_j)\neq ((M+1,2k-1),(M+1,2k))$ for each $k$},\\
 1/z_2^* & \textrm{if $x_j-y_j=(0,\pm 1)$ and $(x_j,y_j)=((M+1,2k-1),(M+1,2k))$ for some $k$}.
  \end{array} \right.
\eea
Clearly, the only terms in the sum that contribute to $Z_{\Lambda^*}(K_1^*,K_2^*,\mathfrak{h})$ are of the form
\bea 
&&z_2^*\sigma_{M+1,1}\sigma_{M+1,2}z_2^*\sigma_{M+1,3}\sigma_{M+1,4}...z_2^*\sigma_{M+1,N-1}\sigma_{M+1,N}\nn
&&[\kappa(x_1,y_1)\sigma_{x_1}\sigma_{y_1}\kappa(x_2,y_2)\sigma_{x_2}\sigma_{y_2}...\kappa(x_n,y_n)\sigma_{x_n}\sigma_{y_{n}}]\nn
&&[\kappa(x_{n+1},y_{n+1})\sigma_{x_{n+1}}\sigma_{y_{n+1}}\kappa(x_{n+2},y_{n+2})\sigma_{x_{n+2}}\sigma_{y_{n+2}}...\kappa(x_{p},y_{p})\sigma_{x_p}\sigma_{y_{p}}]\nn
&&...[\kappa(x_{q+1},y_{q+1})\sigma_{x_{q+1}}\sigma_{y_{q+1}}\kappa(x_{q+2},y_{q+2})\sigma_{x_{q+2}}\sigma_{y_{q+2}}...\kappa(x_{r},y_{r})\sigma_{x_r}\sigma_{y_{r}}] \nonumber 
\eea
or else of the form
\bea 
&&z_2^*\sigma_{M+1,2}\sigma_{M+1,3}z_2^*\sigma_{M+1,4}\sigma_{M+1,5}...z_2^*\sigma_{M+1,N}\sigma_{M+1,1}\nn
&&[\kappa(x_1,y_1)\sigma_{x_1}\sigma_{y_1}\kappa(x_2,y_2)\sigma_{x_2}\sigma_{y_2}...\kappa(x_n,y_n)\sigma_{x_n}\sigma_{y_{n}}]\nn
&&[\kappa(x_{n+1},y_{n+1})\sigma_{x_{n+1}}\sigma_{y_{n+1}}\kappa(x_{n+2},y_{n+2})\sigma_{x_{n+2}}\sigma_{y_{n+2}}...\kappa(x_{p},y_{p})\sigma_{x_p}\sigma_{y_{p}}]\nn
&&...[\kappa(x_{q+1},y_{q+1})\sigma_{x_{q+1}}\sigma_{y_{q+1}}\kappa(x_{q+2},y_{q+2})\sigma_{x_{q+2}}\sigma_{y_{q+2}}...\kappa(x_{r},y_{r})\sigma_{x_r}\sigma_{y_{r}}] \nonumber 
\eea
where $x_j$ and $y_{j}$ are nearest neighbors, $y_j=x_{j+1}$ and $x_1=y_n$, $x_{n+1}=y_p$,..., $x_{q+1}=y_r$. Let each pair $(x_j,y_j)$ be marked by a dimer of weight $\kappa(x_j,y_j)$. Should two dimers have the same position, then clearly one will have weight $z_2^*$ and one $1/z_2^*$, and thus they will form one dimer of weight 1; that is, no dimer at all. Let a contributing path as above be called $\gamma$, and let $||\gamma||:=(z_1^*)^{v}(z_2^*)^{h}$ where $v$ is the number of vertical dimers and $h$ is the number of horizontal dimers. It follows that $Z_{\Lambda^*}(K_1^*,K_2^*,\mathfrak{h})$ can be written as
\bea && Z_{\Lambda^*}(K_1^*,K_2^*,\mathfrak{h})=2^{(M+1)N}(\cosh{K_1^*})^{MN}(\cosh{K_2^*})^{(M+1)N} \sum_{\gamma}||\gamma||
\label{dpf4}
\eea
where the sum is over all contributing paths $\gamma$. Since $\tanh{K_l^*}=e^{-2K_l}$, it follows that
\be \sum_{\gamma}||\gamma||=2\sum_{\lambda}e^{-2|\lambda|/kT},
\ee
and hence the lemma follows.

{\bf Remark} The general argument concerning duality relations was first given in ref. \cite{B}.

McCoy and Wu did not restrict the magnetic field on the boundary, $\mathfrak{H}$, 
to any particular value, so the calculation done in ref. \cite{MW} is more general than necessary to solve our particular problem. What follows is a proof of (\ref{p}).

It follows from (\ref{hi}) and (\ref{pf}) that $Z_{\Lambda}(K_1,K_2,0)$ is
\bea Z_{\Lambda}(K_1,K_2,0)&=&(\cosh{K_1})^{MN}(\cosh{K_2})^{(M+1)N}\nn
&&\sum_{\sigma \in \{-1,1\}^{\Lambda}}\bigg(\prod_{j=1}^M\prod_{k=1}^N(1+\sigma_{j,k}\sigma_{j,k+1}z_1)(1+\sigma_{j,k}\sigma_{j+1,k}z_2)\bigg)\prod_{l=1}^N(1+\sigma_{1,l}z_2)
\label{z0}
\eea
where $z_l=\tanh{K_l}$. In the same way it can be shown \cite{B} that
\bea Z_{\Lambda^*}(K_1^*,K_2^*,\mathfrak{h})&=&(\cosh{K_1^*})^{MN}(\cosh{K_2^*})^{(M+1)N}\nn
&&\sum_{\sigma \in \{-1,1\}^{\Lambda^*}}\bigg(\prod_{j=1}^M\prod_{k=1}^N(1+\sigma_{j,k}\sigma_{j+1,k}z_1^*)\bigg)\bigg(\prod_{m=1}^{M+1}\prod_{n=1}^N(1+\sigma_{m,n}\sigma_{m,n+1}z_2^*)\bigg)\prod_{l=1}^Ni\sigma_{M+1,l}.
\label{z01}
\eea

It remains to show that (\ref{p}) holds.  The calculation of the partition function $Z_{\Lambda^{*}}(K_1^*,K_2^*,\mathfrak{h})$ is just a special case of a calculation already done by McCoy and Wu {\cite{MW}}. Since $\sinh{i\pi/2}=i$, it follows from equation (3.26) of ref. \cite{MW} that
\bea Z_{\Lambda^{*}}(K_1^*,K_2^*,\mathfrak{h})&=&2^{(M+1)N}(\cosh{K_2^*})^{(M+1)N}(\cosh{K_1^*})^{MN}{z_1^*}^{MN/2}(1-{z_2^*}^2)^{(M+1)N/2}\nn
&&\prod_{j=1}^{N/2}\frac{|1+e^{i\theta}|^2}{2\sin{\theta_j}}\mathfrak{v}_j{\mathfrak{v}'_j}({\alpha_j}^{M+1}-{\alpha_j}^{-(M+1)})\nn
&=&2^{(M+1)N}(\cosh{K_2^*})^{(M+1)N}(\cosh{K_1^*})^{MN}{z_1^*}^{MN/2}(1-{z_2^*}^2)^{(M+1)N/2}\nn
&&\prod_{j=1}^{N/2}\frac{|1+e^{i\theta}|^2}{2\sin{\theta_j}}\mathfrak{v}_j{\mathfrak{v}_j'}(\alpha_j-{\alpha_j}^{-1})\prod_{k=1}^{M}(\alpha_j-2 \cos{\varphi_k}+{\alpha_j}^{-1}),
\label{ZMW}\eea
where $\alpha_j$ is the larger root of the quadratic equation
\be (1+{z_1^*}^2)(1+{z_2^*}^2)-2z_2^*(1-{z_1^*}^2)\cos{\theta_j}-z_1^*(1-{z_2^*}^2)(\alpha+\alpha^{-1})=0.
\label{aeqn0}
\ee
Explicitly
\bea
\alpha_j=\frac{1}{2z_1^*(1-{z_2^*}^2)}&&\{(1+{z_1^*}^2)(1+{z_2^*}^2)-2z_2^*(1-{z_1^*}^2)\cos{\theta_j}\nn
&&+(1-{z_1^*}^2)[(1-\hat{\alpha}_1e^{i\theta_j})(1-\hat{\alpha}_1e^{-i\theta_j})(1-\hat{\alpha}_2^{-1}e^{i\theta_j})(1-\hat{\alpha}_2^{-1}e^{-i\theta_j})]^{1/2}\},
\eea
where
\bea
\hat{\alpha}_1&=&z_2^*(1-z_1^*)/(1+z_1^*),\nn
\hat{\alpha}_2&=&{z_2^*}^{-1}(1-z_1^*)/(1+z_1^*).
\eea
$\mathfrak{v}_j$ and $\mathfrak{v}'_j$ are given by
\be \mathfrak{v}_j=\left(\frac{1}{2}\left(1-\frac{{z_1^*}^2|1+z_2^*e^{i\theta_j}|^2-4{z_2^*}^2\sin^2{\theta_j}|1+z_2^*e^{i\theta_j}|^{-2}-(1-{z_2^*}^2)^2|1+z_2^*e^{i\theta_j}|^{-2}}{(1-{z_1^*}^2)[(1-\hat{\alpha}_1e^{i\theta_j})(1-\hat{\alpha}_1e^{-i\theta_j})(1-\hat{\alpha}_2^{-1}e^{i\theta_j})(1-\hat{\alpha}_2^{-1}e^{-i\theta_j})]^{1/2}}\right)\right)^{1/2}
\ee
and
\be \mathfrak{v}'_j=\left(\frac{1}{2}\left(1+\frac{{z_1^*}^2|1+z_2^*e^{i\theta_j}|^2-4{z_2^*}^2\sin^2{\theta_j}|1+z_2^*e^{i\theta_j}|^{-2}-(1-{z_2^*}^2)^2|1+z_2^*e^{i\theta_j}|^{-2}}{(1-{z_1^*}^2)[(1-\hat{\alpha}_1e^{i\theta_j})(1-\hat{\alpha}_1e^{-i\theta_j})(1+\hat{\alpha}_2^{-1}e^{i\theta_j})(1-\hat{\alpha}_2^{-1}e^{-i\theta_j})]^{1/2}}\right)\right)^{1/2}.
\ee
Since 
\be \alpha_j-\alpha_j^{-1}=\frac{1-{z_1^*}^2}{z_1^*(1-{z_2^*}^2)}[(1-\hat{\alpha}_1e^{i\theta_j})(1-\hat{\alpha}_1e^{-i\theta_j})(1-\hat{\alpha}_2^{-1}e^{i\theta_j})(1-\hat{\alpha}_2^{-1}e^{-i\theta_j})]^{1/2},\ee 
it follows from (\ref{ZMW}) that 
\bea Z_{\Lambda^{*}}(K_1^*,K_2^*,\mathfrak{h})&=&2^{(M+1)N}(\cosh{K_2^*})^{(M+1)N}(\cosh{K_1^*})^{MN}\left(\frac{1-{z_1^*}^2}{z_1^*}\right)^{N/2}\nn
\prod_{j=1}^{N/2}&&\frac{|1+e^{i\theta_j}|^2}{2\sin{\theta_j}}\mathfrak{v}_j^*{\mathfrak{v}_j'}^*[(1-\hat{\alpha}_1^*e^{i\theta_j})(1-\hat{\alpha}_1^*e^{-i\theta_j})(1-e^{i\theta_j}/\hat{\alpha}_2^*)(1-e^{-i\theta_j}/\hat{\alpha}_2^*)]^{1/2}\nn
\prod_{k=1}^{M}&&\{(1+{z_1^*}^2)(1+{z_2^*}^2)-2z_2^*(1-{z_1^*}^2)\cos{\theta_j}-2z_1^*(1-{z_2^*}^2)\cos{\varphi_k}\}.
\label{Z2}
\eea
Since 
\bea &&\cosh{2K_l^*}/\cosh{2K_l}=\sinh{2K_l^*},\nn 
&&\prod_{j=1}^{N/2}\frac{|1+e^{i\theta_j}|^2}{2\sin{\theta_j}}=1 \nn
\textrm{and}&& \prod_{j=1}^{N/2}\mathfrak{v}_j{\mathfrak{v}_j'}[(1-\hat{\alpha}_1e^{i\theta_j})(1-\hat{\alpha}_1e^{-i\theta_j})(1-e^{i\theta_j}/\hat{\alpha}_2)(1-e^{-i\theta_j}/\hat{\alpha}_2)]^{1/2}=2(1-{z_1^*}^2)^{-N/2}(z_1^*z_2^*)^{N/2},
\label{eq0}
\eea
it follows from (\ref{Z2}) that
\bea Z_{\Lambda^{*}}(K_1^*,K_2^*,\mathfrak{h})&=&2^{(M+1)N+1}(\cosh{K_2^*})^{(M+1)N}(\cosh{K_1^*})^{MN}{z_2^*}^{N/2}\nn
&&\prod_{j=1}^{N/2}\prod_{k=1}^{M}\{(1+{z_1^*}^2)(1+{z_2^*}^2)-2z_2^*(1-{z_1^*}^2)\cos{\theta_j}-2z_1^*(1-{z_2^*}^2)\cos{\varphi_k}\}\nn
&=&2^{MN+N/2+1}(\sinh{2K_1^*})^{MN/2}(\sinh{2K_2^*})^{(M+1)N/2}\nn
&&\prod_{j=1}^{N/2}\prod_{k=1}^{M}\{\cosh{2K_1}\cosh{2K_2}-\sinh{2K_1}\cos{\theta_j}-\sinh{2K_2}\cos{\varphi_k}\}.
\label{Z3}\eea
It follows from (\ref{dr0}), (\ref{Zb}) and (\ref{Z3}) that (\ref{p}) holds.

\section{The calculation of $Z_{\Lambda}(K_1,K_2,i\pi/2)$}
\label{bkm}

\begin{figure}[bt]
{\epsfig{file=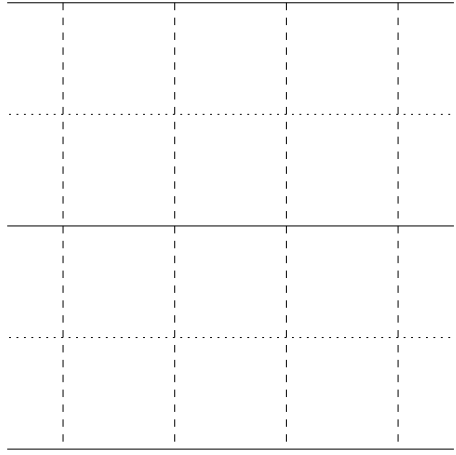, width=6cm}}
\caption{The interactions on the dual lattice $\Lambda^{*}$. Here $M=N=4$. This figure shows the three kinds of bond. The bond $K_1^*$ is shown as a broken line, $K_2^*$ is shown as a solid line, and $-K_2^*$ is shown as a dotted line. There is a magnetic field $\mathfrak{H}^*/kT^*=i\pi/2$ on the lower boundary.}
\label{fig4}
\end{figure}

The partition function in this case can be found from (\ref{hi}) and (\ref{pf}) to be
\bea Z_{\Lambda}(K_1,K_2,i\pi/2)
&=&(\cosh{K_1})^{MN}(\cosh{K_2})^{(M+1)N}\nn
&&\sum_{\sigma \in \{-1,1\}^{\Lambda}}\bigg(\prod_{j=1}^M\prod_{k=1}^Ni\sigma_{j,k}(1+\sigma_{j,k}\sigma_{j,k+1}z_1)(1+\sigma_{j,k}\sigma_{j+1,k}z_2)\bigg)\prod_{l=1}^N(1+\sigma_{1,l}z_2)
\label{z1}
\eea
where $z_l=\tanh{K_l}$. This can be written as 
\bea Z_{\Lambda}(K_1,K_2,i\pi/2)&=&(\cosh{K_1})^{MN}(\cosh{K_2})^{(M+1)N}z_2^{MN/2}\nn
&&\sum_{\sigma \in \{-1,1\}^{\Lambda}}\bigg(\prod_{j=1}^M\prod_{k=1}^N(1+\sigma_{j,k}\sigma_{j,k+1}z_1)(1+\sigma_{j,k}\sigma_{j+1,k}\hat{z}^{(j)}_2)\bigg)\prod_{l=1}^N(1+\sigma_{1,l}z_2)
\label{z2}
\eea
where 
\bea
\hat{z}^{(j)}_2 = \left\{ \begin{array}{ll}
 z_2 & \textrm{if $j$ is even,}\\
 1/z_2 & \textrm{if $j$ is odd.}
  \end{array} \right.
\label{hat}
\eea
(\ref{hat}) gives interactions $K_1$, $K_2$ and $\hat{K}_2$ on $\Lambda$, with the definition of $\hat{K}_2$ being
\be \tanh{\hat{K}_2}=1/\tanh{K_2}.
\label{Khat}
\ee
This defines a new Hamiltonian
\bea \hat{\mathcal{E}}_{\Lambda}(\sigma,E_1,E_2,\hat{E}_2,H)&=&-E_1\sum_{j=1}^M\sum_{k=1}^N\sigma_{j,k}\sigma_{j,k+1}
-E_2\sum_{l=0}^{M/2}\sum_{k=1}^N\sigma_{2l,k}\sigma_{2l+1,k}\nn
&&-\hat{E}_2\sum_{l=0}^{M/2-1}\sum_{k=1}^N\sigma_{2l+1,k}\sigma_{2l+2,k}
-\sum_{j=1}^M\sum_{k=1}^NH(j,k)\sigma_{j,k},
\label{hn}
\eea
and a new partition function
\be \hat{Z}_{\Lambda}(K_1,K_2,\hat{K}_2,h)=\sum_{\sigma \in \{-1,1\}^{\Lambda}}\exp{-\hat{\mathcal{E}}_{\Lambda}(\sigma,E_1,E_2,\hat{E}_2,H)/kT}.
\label{npf}
\ee
In this way, (\ref{z2}) can be written as
\be Z_{\Lambda}(K_1,K_2,i\pi/2)=\frac{(\sinh{K_2})^{MN/2}}{(\cosh{\hat{K}_2})^{MN/2}}\hat{Z}_{\Lambda}(K_1,K_2,\hat{K}_2,0).
\label{rel}
\ee
The Hamiltonian on $\Lambda^*$ is
\bea \hat{\mathcal{E}}_{\Lambda^*}(\sigma,E_1^*,E_2^*,\hat{E}_2^*,H^*)&=&-E_1^*\sum_{j=1}^{M+1}\sum_{k=1}^N\sigma_{j,k}\sigma_{j+1,k}
-E_2^*\sum_{l=0}^{M/2}\sum_{k=1}^N\sigma_{2l,k}\sigma_{2l,k+1}\nn
&&-{\hat{E}}^*_2\sum_{l=0}^{M/2-1}\sum_{k=1}^N\sigma_{2l+1,k}\sigma_{2l+1,k+1}
-\sum_{j=1}^{M+1}\sum_{k=1}^NH^*(j,k)\sigma_{j,k},
\label{hnn}
\eea
and its corresponding partition function may be defined as
\be \hat{Z}_{\Lambda^*}(K_1^*,K_2^*,\hat{K}_2^*,h^*)=\sum_{\sigma \in \{-1,1\}^{\Lambda^*}}\exp{-\hat{\mathcal{E}}_{\Lambda^*}(\sigma,E_1^*,E_2^*,\hat{E}_2^*,H^*)/kT^*},
\label{ndpf}
\ee
where $\Lambda^*$ is the dual lattice (see fig. \ref{fig4}) and 
\be \sinh{2K_1}\sinh{2K_1^*}=\sinh{2K_2}\sinh{2K_2^*}=\sinh{2\hat{K}_2}\sinh{2\hat{K}_2^*}=1.
\label{dr}
\ee
In the same way as (\ref{Zb}) was obtained, one obtains the relation
\bea \hat{Z}_{\Lambda}(K_1,K_2,\hat{K}_2,0)&=&2^{MN-1}(\cosh{K_1})^{MN}(\cosh{K_2})^{(M+1)N}\nn
&&[(\tanh{K_1})^{MN}(\tanh{K_2})^{(M/2+1)N}(\tanh{\hat{K}_2})^{MN/2}]^{1/2}\hat{Z}_{\Lambda^*}(K_1^*,K_2^*,\hat{K}_2^*,\mathfrak{h}).
\label{zs}
\eea
It follows from (\ref{Khat}) and (\ref{dr}) that 
\be \tanh{\hat{K}_2^*}=-\tanh{K_2^*}.
\label{hd}
\ee
Therefore the lattice constant $\hat{K}_2^*$ on $\Lambda^*$ may be taken to be such that
\be \hat{K}_2^*=-K_2^*.
\label{d3}
\ee
Hence (\ref{zs}) simplifies to become
\bea \hat{Z}_{\Lambda}(K_1,K_2,\hat{K}_2,0)&=&2^{MN/2-N/2-1}(\sinh{2K_1})^{MN/2}
(\sinh{2K_2})^{N/2}
(\cosh{K_2})^{MN}
\hat{Z}_{\Lambda^*}(K_1^*,K_2^*,\hat{K}_2^*,\mathfrak{h}).
\label{zs2}
\eea
$Z_{\Lambda^*}(K_1^*,K_2^*,\hat{K}_2^*,\mathfrak{h})$ is given by
\bea Z_{\Lambda^*}(K_1^*,K_2^*,\hat{K}_2^*,\mathfrak{h})&=&(\cosh{K_1^*})^{MN}(\cosh{K_2^*})^{(M+1)N}\nn
\sum_{\sigma \in \{-1,1\}^{\Lambda^*}}&&\bigg(\prod_{j=1}^M\prod_{k=1}^N(1+\sigma_{j,k}\sigma_{j+1,k}z_1^*)\bigg)\bigg(\prod_{m=1}^{M+1}\prod_{n=1}^N(1+\sigma_{m,n}\sigma_{m,n+1}{\hat{z}_2}^{*(m)})\bigg)\prod_{l=1}^Ni\sigma_{M+1,l}.
\label{z3}
\eea
where ${\hat{z}_2}^{*(m)}=(-1)^{m+1}{z_2}^{*}$. Let $\epsilon >0$, and let 
\be h_{\epsilon}=i\pi/2+\epsilon ~{\rm and}~ z_{\epsilon}=\tanh{h_{\epsilon}}.
\ee
The product $\prod_{l=1}^Ni\sigma_{M+1,l}$ can be written as
\be \prod_{l=1}^Ni\sigma_{M+1,l}=\lim_{\epsilon \rightarrow 0}(\cosh{h_{\epsilon}})^N\prod_{l=1}^N(1+\sigma_{M+1,l}z_{\epsilon}).
\label{pr}
\ee

As shown in ref. \cite{mpw}, the sum $\sum$ of (\ref{z3}) can be written as 
\be \sum =2^{(M+1)N-1}\lim_{\epsilon \rightarrow 0}(\cosh{h_{\epsilon}})^N{\rm Pf}A_{\epsilon},
\label{s}
\ee
where $A_{\epsilon}$ is a matrix which will be given below, and ${\rm Pf}{A_{\epsilon}}$ is the Pfaffian of $A_{\epsilon}$. The determinant is given by \cite{MW} \cite{mwb}
\be \det{A_{\epsilon}}=\prod_{j=1}^{N}\det{\mathfrak{B}^{(\epsilon)}(\theta_j)}
\label{da}
\ee
where $\mathfrak{B}^{(\epsilon)}(\theta_j)$ is a $4(M+2)\times 4(M+2)$ matrix given by
\bea
\mathfrak{B}^{(\epsilon)}_{m,m}(\theta_j) =
\left( \begin{array}{cccc}
0 & 1+(-1)^{m+1}z_2^*e^{i\theta_j} & -1 & -1\\
-1-(-1)^{m+1}z_2^*e^{-i\theta_j} & 0 & 1 & -1\\
1 & -1 & 0 & 1\\
1 & 1 & -1 & 0
\end{array} \right)
\label{bjj}
\eea
for $1\leq m\leq M+1$,
\bea
\mathfrak{B}^{(\epsilon)}_{0,0}(\theta_j) =
\left( \begin{array}{cccc}
0 & 1+e^{i\theta_j} & -1 & -1\\
-1-e^{-i\theta_j} & 0 & 1 & -1\\
1 & -1 & 0 & 1\\
1 & 1 & -1 & 0
\end{array} \right),
\label{b00}
\eea
\bea
\mathfrak{B}^{(\epsilon)}_{m,m+1}(\theta_j)=-\mathfrak{B}^{(\epsilon)}_{m+1,m}(\theta_j)^T=
\left( \begin{array}{cccc}
0 & 0 & 0 & 0\\
0 & 0 & 0 & 0\\
0 & 0 & 0 & z_1^*\\
0 & 0 & 0 & 0
\end{array} \right)
\label{bm1}
\eea
for $1\leq m\leq M$, and
\bea
\mathfrak{B}^{(\epsilon)}_{0,1}(\theta_j)=-\mathfrak{B}^{(\epsilon)}_{1,0}(\theta_j)^T=
\left( \begin{array}{cccc}
0 & 0 & 0 & 0\\
0 & 0 & 0 & 0\\
0 & 0 & 0 & z_{\epsilon}\\
0 & 0 & 0 & 0
\end{array} \right).
\label{b01}
\eea
All other elements are zero. 
Let 
\bea \mathfrak{a}_{\pm}&=&\pm 2iz_2^*\sin{\theta_j}|1\pm z_2^*e^{i\theta_j}|^{-2},\nn
\mathfrak{b}_{\pm}&=&\pm (1-{z_2^*}^2)|1\pm z_2^*e^{i\theta_j}|^{-2},\nn
\mathfrak{c}&=&2i\sin{\theta_j}|1+e^{i\theta_j}|^{-2}.
\label{abc}
\eea
Using the same procedure as in ref. \cite{MW}, one finds that $\det{A}$ is given by 
\be \det{A_{\epsilon}}=\prod_{j=1}^N|1+e^{i\theta_j}|^2|1+z_2^*e^{i\theta_j}|^{M+2}|1-z_2^*e^{i\theta_j}|^{M}\det{\mathfrak{C^{(\epsilon)}}(\theta_j)},
\label{deta}
\ee
where
\bea
\mathfrak{C^{(\epsilon)}}(\theta_j) =
\left( \begin{array}{ccccccccccc}
-\mathfrak{c} & 0 & & & & & & & & &\\
0 & \mathfrak{c} & z_{\epsilon} & & & & & & & &\\
 & -z_{\epsilon} & -\mathfrak{a}_+ & \mathfrak{b}_+ & & & & & & &\\ 
 & & -\mathfrak{b}_+ & \mathfrak{a}_+ & z_1^* & & & & & &\\
 & & & -z_1^* & -\mathfrak{a}_- & \mathfrak{b}_- & & & & &\\
 & & & & -\mathfrak{b}_- & \mathfrak{a}_- & z_1^* & & & &\\
 & & & & & -z_1^* & & & & &\\
 & & & & & & & \ddots & & &\\ 
 & & & & & & & & & -\mathfrak{a}_+ & \mathfrak{b}_+\\
 & & & & & & & & & -\mathfrak{b}_+ & \mathfrak{a}_+
\end{array} \right).
\label{cmat}
\eea
The dimension of $\mathfrak{C^{(\epsilon)}}$ is $2(M+2)\times 2(M+2)$. Let $\mathfrak{C}^{(\epsilon)}_{n}$ be the determinant of the $2(n+1)\times 2(n+1)$ matrix of the form (\ref{cmat}), and let $\mathfrak{D}^{(\epsilon)}_{n}$ be the $(2n+1)\times (2n+1)$ determinant with the last row and the last column removed. Then
\bea
\left( \begin{array}{c}
\mathfrak{C}^{(\epsilon)}_{2m+1}\\
z_1^*\mathfrak{D}^{(\epsilon)}_{2m+1}
\end{array} \right)=
\left( \begin{array}{cc}
-\mathfrak{a}_+^2+\mathfrak{b}_+^2 & \mathfrak{a}_+z_1^*\\
-\mathfrak{a}_+z_1^* & {z_1^*}^2
\end{array} \right)
\left( \begin{array}{cc}
-\mathfrak{a}_-^2+\mathfrak{b}_-^2 & \mathfrak{a}_-z_1^*\\
-\mathfrak{a}_-z_1^* & {z_1^*}^2
\end{array} \right)
\left( \begin{array}{c}
\mathfrak{C}^{(\epsilon)}_{2m-1}\\
z_1^*\mathfrak{D}^{(\epsilon)}_{2m-1}
\end{array} \right);~1\leq m\leq M/2
\label{pmat}
\eea
and
\bea
\left( \begin{array}{c}
\mathfrak{C}^{(\epsilon)}_{1}\\
z_1^*\mathfrak{D}^{(\epsilon)}_{1}
\end{array} \right)=
\left( \begin{array}{cc}
-\mathfrak{a}_+^2+\mathfrak{b}_+^2 & \mathfrak{a}_+z_1^*\\
-\mathfrak{a}_+z_1^* & {z_1^*}^2
\end{array} \right)
\left( \begin{array}{c}
\mathfrak{C}_{0}\\
z_{\epsilon}^2{z_1^*}^{-1}\mathfrak{D}_{0}
\end{array} \right),
\label{pmat2}
\eea
where
\be \mathfrak{C}_{0}=-\mathfrak{c}^2 ~{\rm and}~ \mathfrak{D}_{0}=-\mathfrak{c}.
\ee
(\ref{pmat2}) becomes 
\bea
\left( \begin{array}{c}
\mathfrak{C}^{(\epsilon)}_{1}\\
z_1^*\mathfrak{D}^{(\epsilon)}_{1}
\end{array} \right)=-z_{\epsilon}^2\mathfrak{c}
\left( \begin{array}{c}
\mathfrak{a}_+\\
z_1^*
\end{array} \right)
+\left( \begin{array}{c}
o(1)\\
o(1)
\end{array} \right).
\label{pmat3}
\eea
The matrix
\bea 
\mathfrak{P}&:=&\left( \begin{array}{cc}
-\mathfrak{a}_+^2+\mathfrak{b}_+^2 & \mathfrak{a}_+z_1^*\\
-\mathfrak{a}_+z_1^* & {z_1^*}^2
\end{array} \right)
\left( \begin{array}{cc}
-\mathfrak{a}_-^2+\mathfrak{b}_-^2 & \mathfrak{a}_-z_1^*\\
-\mathfrak{a}_-z_1^* & {z_1^*}^2
\end{array} \right)\nn
&=&\left( \begin{array}{cc}
1-4{z_1^*}^2{z_2^*}^2\sin^2{\theta_j}|1-{z_2^*}^2e^{i2\theta_j}|^{-2} & 2iz_1^*(1-{z_1^*}^2)z_2^*\sin{\theta_j}|1-z_2^*e^{i\theta_j}|^{-2}\\
2iz_1^*(1-{z_1^*}^2)z_2^*\sin{\theta_j}|1+z_2^*e^{i\theta_j}|^{-2} & {z_1^*}^4-4{z_1^*}^2{z_2^*}^2\sin^2{\theta_j}|1-{z_2^*}^2e^{i2\theta_j}|^{-2}
\end{array} \right)
\label{pmat4}
\eea
has eigenvalues 
\bea \lambda&=&{z_1^*}^2(1-{z_2^*}^2)^2\alpha/|1-{z_2^*}^2e^{i2\theta_{j}}|^2\nn
{\rm and}~\lambda'&=&{z_1^*}^2(1-{z_2^*}^2)^2/\alpha|1-{z_2^*}^2e^{i2\theta_{j}}|^2
\label{eval}
\eea
where $\alpha$ is the larger root of the quadratic equation 
\be (1+{z_1^*}^4)(1+{z_2^*}^4)-4{z_1^*}^2{z_2^*}^2-2{z_2^*}^2(1-{z_1^*}^2)^2\cos{2\theta_j}-{z_1^*}^2(1-{z_2^*}^2)^2(\alpha+1/\alpha)=0.
\label{aeqn}
\ee
$\mathfrak{P}$ has corresponding eigenvectors $v_1$ and $v_2$. The eigenvector with eigenvalue $\lambda$ is
\bea v_1=\left( \begin{array}{c}
a_1\\
b_1
\end{array} \right)
=a_1\left( \begin{array}{c}
1\\
-\frac{|1+z_2^*e^{i\theta_j}|^2}{2iz_1^*z_2^*(1-{z_1^*}^2)\sin{\theta_j}}\Big(1-\frac{4{z_1^*}^2{z_2^*}^2\sin^2{\theta_j}}{|1-{z_2^*}^2e^{i2\theta_j}|^2}-\frac{4{z_1^*}^2(1-{z_2^*}^2)^2\alpha}{|1-{z_2^*}^2e^{i2\theta_j}|^2}\Big)
\end{array} \right)
\label{evec1}
\eea
and the eigenvector with eigenvalue $\lambda'$ is
\bea v_2=\left( \begin{array}{c}
a_2\\
b_2
\end{array} \right)
=a_2\left( \begin{array}{c}
1\\
-\frac{|1+z_2^*e^{i\theta_j}|^2}{2iz_1^*z_2^*(1-{z_1^*}^2)\sin{\theta_j}}\Big(1-\frac{4{z_1^*}^2{z_2^*}^2\sin^2{\theta_j}}{|1-{z_2^*}^2e^{i2\theta_j}|^2}-\frac{4{z_1^*}^2(1-{z_2^*}^2)^2}{|1-{z_2^*}^2e^{i2\theta_j}|^2\alpha}\Big)
\end{array} \right).
\label{evec2}
\eea
It follows from (\ref{pmat}), (\ref{pmat3}) and (\ref{pmat4}) that 
\bea
\left( \begin{array}{c}
\mathfrak{C}^{(\epsilon)}_{M+1}\\
z_1^*\mathfrak{D}^{(\epsilon)}_{M+1}
\end{array} \right)=-z_{\epsilon}^2\mathfrak{c}\mathfrak{P}^{M/2}
\left( \begin{array}{c}
\mathfrak{a}_+\\
z_1^*
\end{array} \right)
+\left( \begin{array}{c}
o(1)\\
o(1)
\end{array} \right).
\label{pmat5}
\eea
Since
\bea \mathfrak{P}^{M/2}=\left( \begin{array}{cc}
a_1 & a_2\\
b_1 & b_2
\end{array} \right)
\left( \begin{array}{cc}
\lambda^{M/2} & 0\\
0 & \lambda'^{M/2}
\end{array} \right)
\left( \begin{array}{cc}
a_1 & a_2\\
b_1 & b_2
\end{array} \right)^{-1},
\label{pmat6}
\eea
it follows that
\bea
\left( \begin{array}{c}
\mathfrak{C}^{(\epsilon)}_{M+1}\\
z_1^*\mathfrak{D}^{(\epsilon)}_{M+1}
\end{array} \right)=-z_{\epsilon}^2\mathfrak{c}\left( \begin{array}{cc}
a_1 & a_2\\
b_1 & b_2
\end{array} \right)
\left( \begin{array}{cc}
\lambda^{M/2} & 0\\
0 & \lambda'^{M/2}
\end{array} \right)
\left( \begin{array}{cc}
a_1 & a_2\\
b_1 & b_2
\end{array} \right)^{-1}
\left( \begin{array}{c}
\mathfrak{a}_+\\
z_1^*
\end{array} \right)
+\left( \begin{array}{c}
o(1)\\
o(1)
\end{array} \right).
\label{pmat7}
\eea
Therefore
\bea \det{\mathfrak{C}^{(\epsilon)}}=\mathfrak{C}^{(\epsilon)}_{M+1}=-\frac{z_{\epsilon}^2\mathfrak{c}}{a_1b_2-a_2b_1}\{a_1\lambda^{M/2}(b_2\mathfrak{a}_+-a_2z_1^*)+a_2\lambda'^{M/2}(-b_1\mathfrak{a}_++a_1z_1^*)\}+o(1).
\label{c}
\eea
Since
\be a_2(-b_1\mathfrak{a}_++a_1z_1^*)=\alpha^{-1}a_1(b_2\mathfrak{a}_+-a_2z_1^*),
\label{eq1}
\ee
(\ref{c}) can be written as 
\bea \det{\mathfrak{C}^{(\epsilon)}}&=&-\frac{z_{\epsilon}^2\mathfrak{c}}{a_1b_2-a_2b_1}a_1(b_2\mathfrak{a}_+-a_2z_1^*)\frac{{z_1^*}^M(1-{z_2^*}^2)^M}{|1-{z_2^*}^2e^{i2\theta_j}|^M}\alpha^{-1/2}(\alpha^{(M+1)/2}+\alpha^{-(M+1)/2})+o(1)\nn
&=&-\frac{z_{\epsilon}^2\mathfrak{c}}{a_1b_2-a_2b_1}a_1(b_2\mathfrak{a}_+-a_2z_1^*)\frac{{z_1^*}^M(1-{z_2^*}^2)^M}{|1-{z_2^*}^2e^{i2\theta_j}|^M}(1+\alpha^{-1})\prod_{k=1}^{M/2}(\alpha+\alpha^{-1}-2\cos{\phi_k})+o(1).
\label{c2}
\eea
Using (\ref{eq1}) again, one obtains 
\be \frac{1}{a_1b_2-a_2b_1}a_1(b_2\mathfrak{a}_+-a_2z_1^*)(1+\alpha^{-1})=\mathfrak{a}_+.
\label{eq2}
\ee
Hence (\ref{c2}) can be written as 
\bea \det{\mathfrak{C}^{(\epsilon)}}=-z_{\epsilon}^2\mathfrak{c}\mathfrak{a}_+\frac{{z_1^*}^M(1-{z_2^*}^2)^M}{|1-{z_2^*}^2e^{i2\theta_j}|^M}\prod_{k=1}^{M/2}(\alpha+\alpha^{-1}-2\cos{\phi_k})+o(1).
\label{c3}
\eea
(\ref{deta}) and (\ref{c3}) give
\bea \det{A_{\epsilon}}=\prod_{j=1}^N|1+z_2^*e^{i\theta_j}|^{2}|1+e^{i\theta_j}|^2z_{\epsilon}^2\mathfrak{c}\mathfrak{a}_+{z_1^*}^M(1-{z_2^*}^2)^M\prod_{k=1}^{M/2}(\alpha+\alpha^{-1}-2\cos{\phi_k})+o(1).
\label{deta2}
\eea
Using (\ref{abc}) and (\ref{aeqn}), (\ref{deta2}) can be written as 
\bea \det{A_{\epsilon}}&=&\prod_{j=1}^N 4z_2^*\sin^2{\theta_j}z_{\epsilon}^2\nn
&&\prod_{k=1}^{M/2}\{(1+{z_1^*}^4)(1+{z_2^*}^4)-4{z_1^*}^2{z_2^*}^2-2{z_2^*}^2(1-{z_1^*}^2)^2\cos{2\theta_j}-2{z_1^*}^2(1-{z_2^*}^2)^2\cos{\phi_k}\}+o(1).
\label{deta3}
\eea
(\ref{def}), (\ref{z3}), (\ref{pr}), (\ref{s}) and (\ref{deta3}) give 
\bea Z_{\Lambda^*}(K_1^*,K_2^*,\hat{K}_2^*,\mathfrak{h})&=&(\cosh{K_1^*})^{MN}(\cosh{K_2^*})^{(M+1)N}\prod_{j=1}^{N/2}4\sin^2{\theta_j}z_2^*\nn
&&\prod_{k=1}^{M/2}\{(1+u_1^2)(1+u_2^2)-4u_1u_2-2u_2(1-u_1)^2\cos{2\theta_j}-2u_1(1-u_2)^2\cos{\phi_k}\}.
\label{deta4}
\eea
Since
\be \prod_{j=1}^{N/2}2\sin{\theta_j}=2,
\label{sp}
\ee
it follows that
\bea \hat{Z}_{\Lambda^*}(K_1^*,K_2^*,\hat{K}_2^*,\mathfrak{h})&=&4(\cosh{K_1^*})^{MN}(\cosh{K_2^*})^{(M+1)N}{z_2^*}^{N/2}\nn
&&\prod_{j=1}^{N/2}\prod_{k=1}^{M/2}\{(1+u_1^2)(1+u_2^2)-4u_1u_2-2u_2(1-u_1)^2\cos{2\theta_j}-2u_1(1-u_2)^2\cos{\phi_k}\}.
\label{deta5}
\eea
It now follows from (\ref{rel}), (\ref{zs2}) and (\ref{s}) that (\ref{bk2}) holds.

\section{Conclusion}
\label{conclusion}
The new results (\ref{p}) and (\ref{bk2}) have been presented.

\acknowledgments
This work was supported by the Belgian Interuniversity Attraction Poles Program P6/02. Equation (\ref{z}) was first found by other means by I. Jensen, J. M. Maillard, B. M. McCoy and R. E. Shrock. This calculation has not yet been published. The author thanks Prof. B. M. McCoy for useful discussions.

\end{document}